\documentclass{ws-p8-50x6-00}

\begin{document}
\title{Meson Interferometry and the Quest for Quark-Gluon Matter}
\author{Sven Soff}
\address{Lawrence Berkeley National Laboratory, One Cyclotron Road, MS70-319, 
Berkeley CA 94720, USA\\
E-mail: ssoff@lbl.gov}
\maketitle
\abstracts{
We point out what we may learn from the investigation 
of identical two-particle interferometry in ultrarelativistic 
heavy ion collisions if  
we assume a particular model scenario by the 
formation of a thermalized quark-gluon plasma hadronizing via 
a first-order phase transition to an interacting hadron gas.  
The bulk properties of the two-pion correlation functions are dominated 
by these late and soft resonance gas rescattering processes. 
However, we show that kaons at large transverse momenta 
have several advantages and a bigger sensitivity to the QCD 
phase transition parameters.}
\vspace*{-0.4cm}
\section{HBT-radius parameters and their relevance for 
the quest for quark-gluon matter}
Correlations of identical particle pairs, also called 
HBT interferometry, provide important information on 
the space-time extension of the particle emitting source as 
for example in ultrarelativistic heavy ion 
collisions. 
In this case, QCD lattice calculations have predicted 
a transition from quark-gluon matter to hadronic matter at high 
temperatures. 
For a first-order phase transition, large hadronization times 
have been expected due to the associated large latent heat as compared to 
a purely hadronic scenario. 
Entropy has to be conserved while the number of degrees of freedom 
is reduced throughout the phase transition. 
Thus, one has expected a considerable jump in the magnitude of 
the HBT-radius parameters and the emission duration 
once the energy density is large enough to produce 
quark-gluon matter\cite{soffbassdumi}. 
The two alternative space-time evolution pictures, 
with and without a phase transition, are 
illustrated in Fig.\ 1 in the $z$-$t$-diagram.   
After the collision of the two nuclei, each  with nucleon number $A$, 
the system is formed at some eigen-time $\tau$ (indicated by the hyperbola) 
and the initial expansion proceeds either in a hadronic state (left-hand side) 
or in a state dominated by partonic degrees of freedom, 
for example a quark-gluon plasma (QGP) (right-hand side). 
In the latter case, the formation of a mixed phase, leads to 
large hadronization times and thus to rather long emission durations. 
The freeze-out is defined as the decoupling of 
the particles, i.e., the space-time coordinates of their last 
(strong) interactions.
As a consequence, HBT interferometry and in particular the 
excitation function of the HBT-parameters have been considered 
as an ideal tool to detect the existence and the properties of a transition 
from a thermalized quark-gluon plasma to hadrons.
\begin{figure}[t]
\epsfxsize=15pc 
\epsfbox{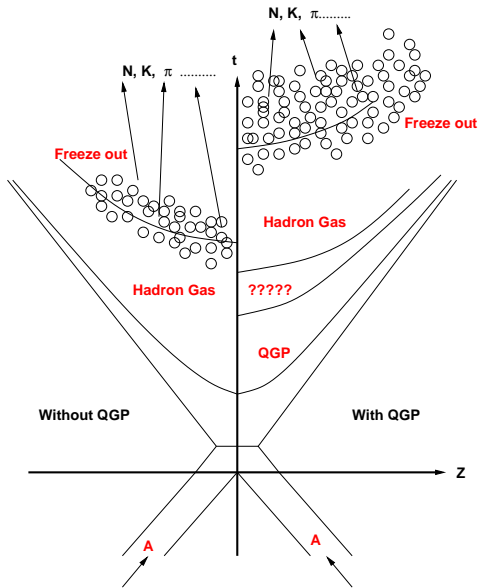} 
\caption{Illustration 
of the space-time evolution in the 
$z$-$t$-diagram with (right) and without (left) a phase 
transition$^2$. 
Proceeding through a first-order phase transition with a 
large latent heat should 
lead to large hadronization times, thus yielding eventually 
large HBT-radius parameters and emission duration. 
\label{fig:minkowski}}
\end{figure}
\vspace*{-0.4cm}
\section{The importance of late soft hadronic rescatterings 
for two-particle correlations at small relative momenta}
Here, we discuss calculations based on a two-phase dynamical transport 
model that describes the early quark-gluon plasma phase 
by hydrodynamics and the later stages after hadronization from 
the phase boundary of the mixed phase by microscopic transport of the 
hadrons \cite{soffbassdumi}. In the hadronic 
phase, resonance (de)excitations and binary collisions are modeled based 
on cross sections and resonance properties as measured in vacuum. 
Fig.~2a shows the pion HBT-parameters $R_i$ as a function of the 
transverse momentum $K_T$ as calculated from the rms-widths of 
the freeze-out distributions \cite{soffbassdumi}. 
$R_{\rm out}$ probes the spatial and temporal extension of the source 
while $R_{\rm side}$ is only sensitive to the spatial extension. 
Thus, the ratio $R_{\rm out}/R_{\rm side}$ gives a measure of the 
emission duration. 
Here, we focus on the fact that for all initial conditions 
considered (SPS or RHIC energies and critical temperatures 
$T_c \simeq 160\,$MeV or $T_c\simeq 200\,$MeV) the HBT-parameters 
appear to be rather similar. 
This demonstrates that a long-lived hadronic phase dominates 
the bulk dependencies of the pion HBT-parameters rather 
than the exact properties of the QCD phase transition. 
\begin{figure}[t]
\epsfxsize=15pc 
\epsfbox{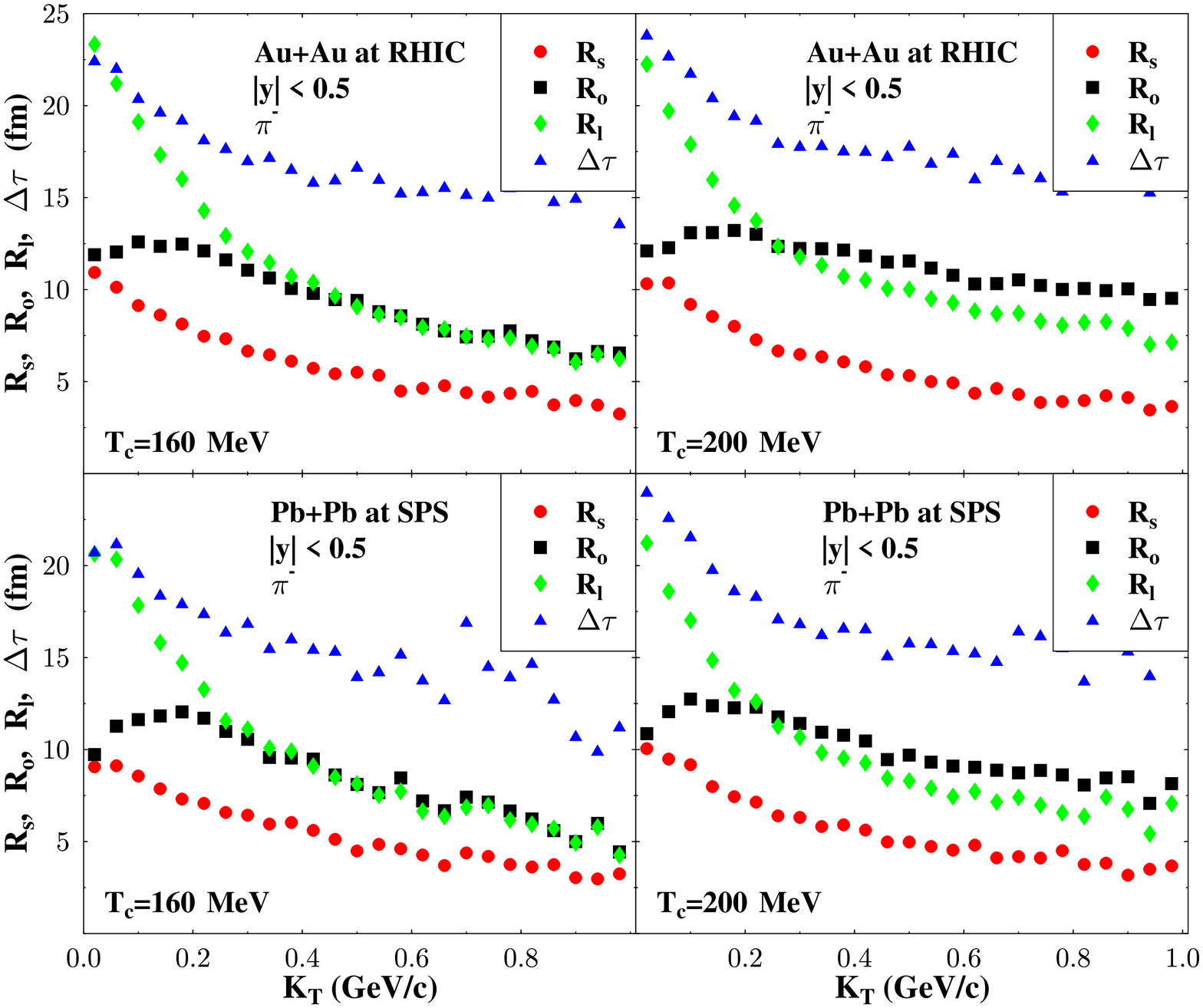}
\epsfxsize=13.5pc
\epsfbox{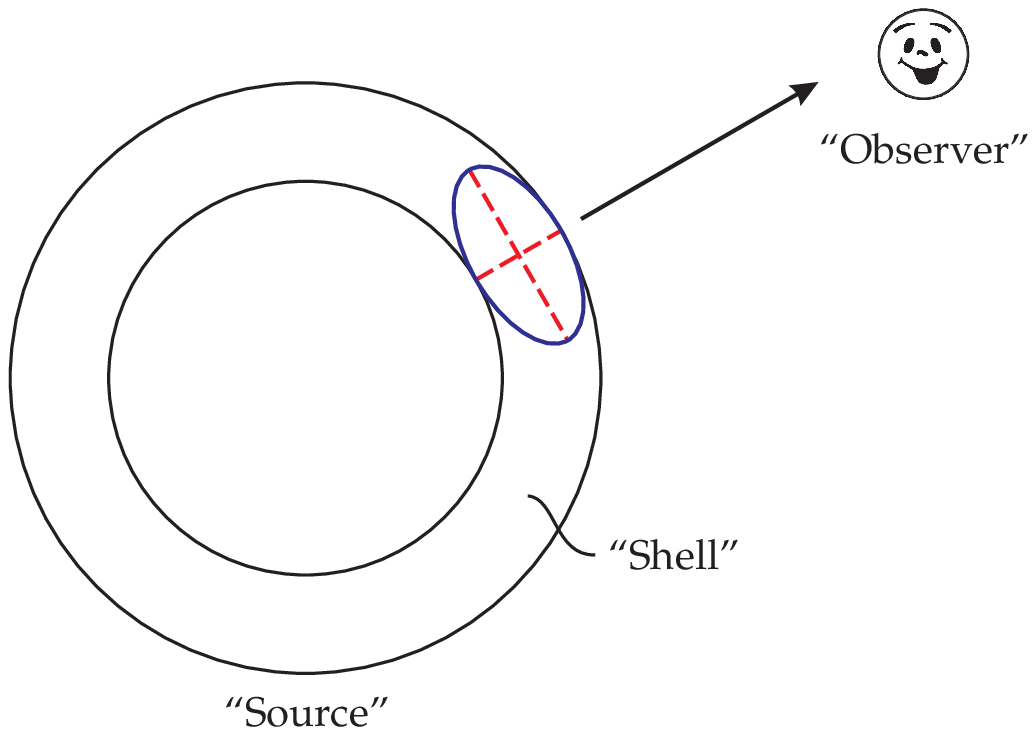} 
\caption{(a) Pion HBT-radius parameters and emission duration as a 
function of $K_T$ as calculated from the rms-radii of 
the {\it QGP+hadronic rescattering} model freeze-out distributions. 
The four different panels correspond to RHIC (top) and SPS (bottom) energies 
and $T_c\simeq 160\,$MeV (left) and $T_c\simeq 200\,$MeV, respectively.
(b) Illustration of a shell-like emission. 
The surface emission geometry corresponds to small values 
of the ratio $R_{\rm out}/R_{\rm side}$, indicated by the two dashed lines 
in the emission volume element relevant for an observer. 
The dashed line in the direction to the observer corresponds to the 
{\it out} direction, the orthogonal line is the {\it side} direction.
\label{fig:rosl}}
\vspace*{-1.0cm}
\end{figure}
In addition, the ratio $R_{\rm out}/R_{\rm side}$ increases 
as a function of $K_T$ up to values of about $1.5-2$ indicating 
the large emission durations. 
However, experimental data at RHIC \cite{STARpreprint,Johnson:2001zi} show a 
completly new behaviour (not seen at SPS). 
The $R_{\rm out}/R_{\rm side}$ ratio decreases and even is smaller than 
unity. This would hint at a rather explosive scenario with 
very short emission times, not compatible with a picture 
of a thermalized quark-gluon plasma hadronizing via a first-order 
phase transition to an interacting hadron gas.
Rather a shell-like emission as illustrated in Fig.2b would be prefered.
Thus, the further study of HBT-interferometry will provide 
extremly important information e.g.\ on the hadronization process or 
the question of thermalization in ultrarelativistic heavy ion collisions.
\vspace*{-0.4cm}
\section{Advantages of Kaons}
Besides many experimental advantages kaons are less contaminated by 
long-lived resonances and escape the opaque hadronic phase easier. 
Thus, $\sim 30\%$ of the kaons at $K_T\sim 1\,$GeV/c are directly emitted 
from the phase-boundary. 
Complementary, large $K_T$ kaons and their $R_{\rm out}/R_{\rm side}$ ratio 
exhibit a strong sensitivity on the QCD equation of state as shown in Fig.~3. 
\begin{figure}[t]
\epsfxsize=15pc 
\epsfbox{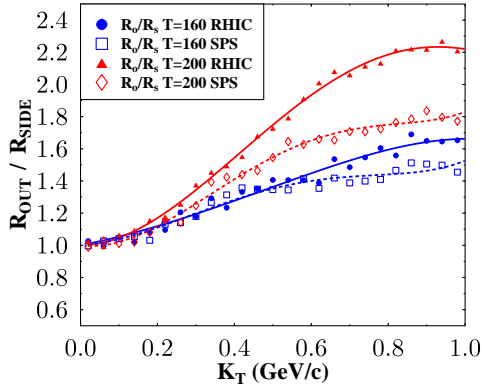} 
\caption{Ratio $R_{\rm out}/R_{\rm side}$ for kaons as a function of 
transverse momentum $K_T$ as calculated with the 
{\it QGP+hadronic rescattering model} for SPS and RHIC and 
critical temperatures $T_c\simeq 160\,$MeV and $T_c \simeq 200\,$MeV. 
\label{fig:kaonrosl}}
\end{figure}
\vspace*{-0.4cm}
\section*{Acknowledgments}
S.S.\ has been supported by the Alexander von Humboldt Foundation through a 
Feodor Lynen Fellowship and 
DOE Grant No.\ DE-AC03-76SF00098.
S.S.\ is grateful to S.~Bass, A.~Dumitru, D.~Hardtke, and S.~Panitkin 
for fruitful collaborations leading to the results presented here.


\vspace*{-0.5cm}
\begin{thebibliography}{99}
\bibitem{soffbassdumi}
S.~Soff, S.~A.~Bass, A.~Dumitru,
{\it Phys.\ Rev.\ Lett.\ }{\bf 86}, 3981 (2001) and references therein;
S.~Soff, S.~Bass, D.~Hardtke, S.~Panitkin,
{\it e-print archive} nucl-th/0109055.
\bibitem{panitkinfig}
Figure drawn and kindly provided by S.~Panitkin.
\bibitem{STARpreprint}
STAR Collaboration, C.~Adler {\it et al.},
{\it Phys.\ Rev.\ Lett.\ }  {\bf 87}, 082301 (2001). 
\bibitem{Johnson:2001zi}
PHENIX Collaboration, S.~C.~Johnson {\it et al.},
nucl-ex/0104020.
\bibitem{Zschiesche:2001dx}
D.~Zschiesche, S.~Schramm, H.~St\"ocker, W.~Greiner,
nucl-th/0107037.
\end{thebibliography}
\end{document}